\documentclass{aastex62}

\usepackage{graphicx}

\graphicspath{{./}{figures/}}

\shorttitle{A new identity card for NGC 6440}
\shortauthors{.}

\begin{document}

\title{A new identity card for the bulge globular cluster NGC
 6440 from resolved star counts\footnote{Based on
    observations collected with the NASA/ESA HST (Prop. 11685, 12517,
    13410, and 15403), obtained at the Space Telescope Science
    Institute, which is operated by AURA, Inc., under NASA contract
    NAS5-26555. Also based on observations collected at the European
    Southern Observatory, Cerro Paranal (Chile), under proposal
    077.D$-$0775(Bale).}}

\correspondingauthor{Cristina Pallanca}
\email{cristina.pallanca3@unibo.it}
\author[0000-0002-7104-2107]{Cristina Pallanca} \affil{Dipartimento di
  Fisica e Astronomia, Universit\`a di Bologna, Via Gobetti 93/2,
  Bologna I-40129, Italy} \affil{Istituto Nazionale di Astrofisica
  (INAF), Osservatorio di Astrofisica e Scienza dello Spazio di
  Bologna, Via Gobetti 93/3, Bologna I-40129, Italy}
\author[0000-0001-5613-4938]{Barbara Lanzoni} \affil{Dipartimento di
  Fisica e Astronomia, Universit\`a di Bologna, Via Gobetti 93/2,
  Bologna I-40129, Italy} \affil{Istituto Nazionale di Astrofisica
  (INAF), Osservatorio di Astrofisica e Scienza dello Spazio di
  Bologna, Via Gobetti 93/3, Bologna I-40129, Italy}
\author[0000-0002-2165-8528]{Francesco R. Ferraro} \affil{Dipartimento
  di Fisica e Astronomia, Universit\`a di Bologna, Via Gobetti 93/2,
  Bologna I-40129, Italy} \affil{Istituto Nazionale di Astrofisica
  (INAF), Osservatorio di Astrofisica e Scienza dello Spazio di
  Bologna, Via Gobetti 93/3, Bologna I-40129, Italy}
\author[0000-0003-2688-7511]{Luca Casagrande} \affil{Research School
  of Astronomy and Astrophysics, The Australian National University,
  Canberra, ACT 2611, Australia} \author[0000-0003-4746-6003]{Sara
  Saracino} \affil{Astrophysics Research Institute, Liverpool John
  Moores University, 146 Brownlow Hill, Liverpool L3 5RF, UK}
\author[0000-0000-0000-0000]{Bhavana Purohith Bhaskar Bhat}
\affil{Dipartimento di Fisica e Astronomia, Universit\`a di Bologna,
  Via Gobetti 93/2, Bologna I-40129, Italy} \affil{Istituto Nazionale
  di Astrofisica (INAF), Osservatorio di Astrofisica e Scienza dello
  Spazio di Bologna, Via Gobetti 93/3, Bologna I-40129, Italy}
\author[0000-0000-0000-0000]{Silvia Leanza} \affil{Dipartimento di
  Fisica e Astronomia, Universit\`a di Bologna, Via Gobetti 93/2,
  Bologna I-40129, Italy} \affil{Istituto Nazionale di Astrofisica
  (INAF), Osservatorio di Astrofisica e Scienza dello Spazio di
  Bologna, Via Gobetti 93/3, Bologna I-40129, Italy}
\author[0000-0003-4237-4601]{Emanuele Dalessandro} \affil{Istituto
  Nazionale di Astrofisica (INAF), Osservatorio di Astrofisica e
  Scienza dello Spazio di Bologna, Via Gobetti 93/3, Bologna I-40129,
  Italy} \author[0000-0003-2742-6872]{Enrico Vesperini}
\affil{Department of Astronomy, Indiana University, Bloomington, IN
  47401, USA}

\begin{abstract}

We present a new identity card for the cluster NGC 6440 in the
Galactic Bulge.  We have used a combination of
high-resolution Hubble Space Telescope images, wide-field ground-based
observations performed with the ESO-FORS2, and the public survey
catalog Pan-STARRS, to determine the gravitational center, projected
density profile and structural parameters of this globular from
resolved star counts.  The new determination of the cluster center
differs by $\sim 2\arcsec$ (corresponding to 0.08 pc) 
from the previous estimate, which was
based on the surface brightness peak. The star density profile,
extending out to $700\arcsec$ from the center and suitably
decontaminated from the Galactic field contribution,
is best-fitted by a King model with significantly larger
  concentration ($c=1.86\pm0.06$) and smaller core radius
  ($r_c=6.4\arcsec\pm0.3$\arcsec) with respect to the literature
  values.   By taking advantage of high-quality optical and
near-infrared color-magnitude diagrams, we also estimated the cluster
age, distance and reddening. The luminosity of the RGB-bump was also
determined. This study indicates that the extinction coefficient in
the  bulge, in the direction of the cluster has a value
($R_V=2.7$) that is significantly smaller than that traditionally used
for the Galaxy ($R_V=3.1$). The corresponding best-fit values of the
age, distance and color excess of NGC 6440 are 13 Gyr, 8.3 kpc and
$E(B-V)\sim 1.27$, respectively.
These new determinations also allowed us to update the values of
  the central ($t_{rc}=2.5\ 10^7$ yr) and half-mass  ($t_{rh}=10^9$ yr) relaxation times, 
  suggesting that NGC 6440 is in a dynamically
  evolved stage.

\end{abstract}


\section{INTRODUCTION}
\label{intro}
We are carrying out the project {\it Cosmic-lab}\footnote{See the web
  page:\texttt{http://www.cosmic-lab.eu/Cosmic-Lab/Home.html}}
aimed at using star clusters in the Local Universe as cosmic
laboratories to study the complex interplay between the dynamical
evolution of stellar systems and the properties of their stellar
populations (e.g., \citealt{ferraro+12, ferraro+18a, ferraro+18b,
  ferraro+19, lanzoni+16, lanzoni+18a, lanzoni+18b, lanzoni+19}).  In
fact,
gravitational interactions among stars have a dramatic effect on the
dynamical evolution of dense stellar systems like globular clusters
(GCs), acting over time scales significantly shorter than their age
and in a way that primarily depends on their internal properties (mass
density, binary content, total mass, etc.) and the external
environment (Galactic tide, disk shocks; e.g. \citealp{bailyn95,
  meylan+97}).  Such interactions have been indicated also as an
efficient channel for the formation of stellar exotica (like
interacting binaries, blue stragglers and millisecond pulsars;
\citealt{bailyn95, pooley+03, ransom+05, ferraro+09a, ferraro+16,
  pallanca10, pallanca13, pallanca14, pallanca17, dalessandro14,
  cadelano17, cadelano18, beccari+19}). Indeed, the innermost core
regions of GCs are expected to offer the ideal environment for the
occurrence of stellar interactions able to generate exotic objects,
including the long sought class of intermediate-mass black holes
\citep[e.g.][]{portegies04, giersz+15}.  For this reason, the
accurate characterization of GCs in terms of structural properties,
dynamical status and internal kinematics is a crucial step to properly
understand how dynamical processes affect the evolutionary history of
these systems and impact the formation of stellar exotica.

The systematic use of Hubble Space Telescope (HST) and, more recently,
adaptive optics assisted observations has opened the possibility of
constructing projected density profiles directly from star counts even
for the innermost regions of high-density stellar systems, not only in
the Galactic halo and bulge (e.g., \citealt{ferraro+09b, saracino+15,
  ferraro+21}), but also beyond the Milky Way \citep{lanzoni+19}.  In
spite of this, the vast majority of GC structural and morphological
parameters currently available in largely used catalogs (e.g.,
\citealt{harris, mackey+03, mclaughlin+05}) are still derived from
surface brightness (SB) profiles. In previous papers (see
\citealt{ferraro+99a, ferraro+03, lanzoni+07a, lanzoni+07b,
  lanzoni+07c,lanzoni+10, lanzoni+19, miocchi+13}), we demonstrated
the advantage of determining the cluster structural parameters from
individual star counts instead of SB.  By construction, SB
  profiles directly depend on the luminosity of the surveyed stars and
  can therefore be artificially distorted by the sparse presence of
  luminous sources \citep[see, e.g.,][for the discussion of methods
    trying to correct for this problem]{ng06} and/or reddening bubbles
  within the field of view (FOV). Instead, this has no impact on the
  density distribution obtained from resolved star counts since every
  object has the same weight independently of its luminosity. In
  addition, if proper motions can be measured, field stars (that may
  be particularly bright and substantially contribute to SB) can be
  explicitly excluded from the determination of the number count
  density profile. Hence, although SB can be helpful in cases of
  severe photometric incompleteness of the catalogs
  \citep[e.g.,][]{santos+20}, it should be always used with caution,
 and number counts generally represent the most robust way for determining
the cluster structural parameters
\citep[e.g.][]{lugger+95,ferraro+99a, ferraro+03}.  Moreover, once the
cluster core is full resolved into individual stars, the determination
of the center of gravity of the system becomes possible by simply
averaging the coordinates of the detected stars. Indeed, we were among
the first teams in promoting and adopting the center of gravity,
instead of the center of luminosity, as optimal proxy of the cluster
center (\citealt{montegriffo+95}, see also \citealt{calzetti+93}).
However, apart from a few studies regarding individual or very small
sets of clusters \citep[e.g.][]{salinas+12}, these techniques have not
be fully exploited in the literature yet, because constructing
complete samples of stars in the highly crowded central regions of GCs
is not an easy task \citep{ferraro+97,ferraro+97b,raso+17}.

Our group already published resolved star density profiles for stellar
systems in different dynamical stages of their evolution (both pre-
and post-core collapse GCs; see \citealp{lanzoni+10, miocchi+13}, and
\citealp{ferraro+03, ferraro+09a,dalessandro+13}, respectively), both
in the Galaxy and in the Large Magellanic Cloud \citep{lanzoni+19}.
Here we present the determination of the star density profile and the
structural parameters for the bulge GC NGC 6440, for which we recently
built a high-resolution extinction map able to correct for the strong
differential reddening effects in the direction of the cluster
\citep{pallanca+19}. The paper is organized as follows. In Section
\ref{data} we describe the used dataset and the main steps of the data
analysis. The procedure adopted to determine the gravity center is
reported in Section \ref{center}.  In Section \ref{dens} we describe
the method used to obtain the observed star count profile and the
determination of the structural parameters through its fit  with
  the King model family. 
Section \ref{DMage} reports the estimate of the distance modulus and
the age.  The identification of the RGB-bump and the comparison with
the literature is presented in Section \ref{bump}. Finally, in Section
\ref{conclu} we discuss the main results and we summarise the
conclusions.

\section{OBSERVATIONS AND DATA ANALYSIS}
\label{data}
To properly sample the entire radial extension of NGC 6440, in the
present work we used a combination of high-resolution and wide-field
images, complemented with catalogs from public surveys.

The highly crowded central regions of the system have been
investigated by means of HST data, consisting in a series of deep
images acquired with the Wide Field Camera 3 (WFC3) through different
filters (especially, F606W and F814W)
and in
various epochs (see Table 1). This is essentially the same dataset
used in \citet{pallanca+19} to construct the extinction map in the
direction of the cluster. Hence more details can be found there.  Here
we just remind that this dataset  provides us with two different
samples: (a) the {\it HST-PM sample} including all the stars with
measured proper motion (PM), thus allowing the decontamination from
possible non-cluster members (Galactic field stars), and (b) the {\it
  HST-noPM sample} including also the stars observed in only one epoch
(hence, with no PM measure) and covering a larger area on the plane of
the sky.
  
For the intermediate cluster region, we used ground-based data
acquired with the FOcal Reducer/low dispersion Spectrograph 2 (FORS2)
mounted at the ESO Very Large Telescope at Paranal Observatory (Chile)
and retrieved from the ESO Science Archive. The FORS2 imaging detector
consists of a mosaic of two 2000x4000 pixel MIT CCDs (15 $\mu$m/pixel)
that combines a relatively large FOV ($6.8\arcmin\times
6.8\arcmin$) and reasonably high-resolution capabilities (pixel size
of $\sim 0.25\arcsec$) for the standard resolution set up.  The core
of the cluster is roughly located at the center of the FORS2 FOV.
Only one image in the $V_{\rm BESS}$ and one in the $I_{\rm BESS}$
filters were available for NGC 6440.

To properly sample the cluster outskirts and beyond, we retrieved
(from {\it https://catalogs.mast.stsci.edu/panstarrs/}) the Pan-STARRS
catalog for a circular region of $700\arcsec$ radius centered on NGC
6440.  The Panoramic Survey Telescope and Rapid Response System
(Pan-STARRS) is a wide-field photometric survey operated by the
Institute for Astronomy at the University of Hawaii, performed with a
1.4 Gigapixel camera (GPC1) mounted at a 1.8 meter telescope, in five
broad-band filters ($g$, $r$, $i$, $z$, $y$). For the present analysis
we used only $i$ and $r$ data.

The detailed description of the data analysis procedure for the HST
dataset can be found in \citet{pallanca+19}.  Here we just summarize
the main steps. The photometric analysis has been carried out by using
the DAOPHOT package \citep{daophot}.  The point spread function (PSF)
for each image has been modelled on several bright and isolated stars,
by using the DAOPHOTII/PSF routine. Then PSF-fitting photometry has
been performed independently on all the images by imposing a source
detection threshold more than 5-$\sigma$ above the background level
and, a master list has been produced considering as reliable sources
all the objects measured in more than half of the images in at least
one filter.  We then run the ALLFRAME package \citep{daophot,allframe}
that simultaneously determines the brightness of the stars in all the
frames, while enforcing one set of centroids and one transformation
between all the images. Finally, the magnitudes obtained for each star
have been normalized to a reference frame and averaged. The
photometric error was derived from the standard deviation of the
repeated measures.  The instrumental magnitudes have been calibrated
to the VEGAMAG system by using the photometric zero points reported on
the WFC3 web
page\footnote{http://www.stsci.edu/hst/wfc3/phot\_zp\_lbn}.  Geometric
distortions have been corrected following the prescription of
\citet{bellini+11} and then reported to the absolute coordinate system
($\alpha,\delta$) as defined by the World Coordinate System by using a
sample of stars in common with the publicly available Gaia DR2 catalog
\citep{gaia16a,gaia16b}. The resulting astrometric accuracy turns out
to be $< 0.1\arcsec$.

A similar procedure was adopted in the analysis of the FORS2
wide-field dataset.  In summary, per each exposure we modelled the PSF
by using dozens of bright, isolated and non saturated stars and we
applied such model to all the sources detected at 3-$\sigma$ above the
background level. In a second step we created a master list containing
all the detected sources and for each object we forced the fit in both
filters. The further step consisted in the creation of the catalog
listing all the magnitudes measured in both filters. Finally, the
instrumental positions have been reported to the absolute coordinate
system by using a sample of stars in common with the Gaia DR2 catalog.
 In order to make the three catalogs homogeneous in magnitude, we
  calibrated the V$_{BESS}$ and I$_{BESS}$ magnitudes of FORS2 onto
  the HST F606W and F814W magnitudes, respectively, by using color
  equations obtained from a large number of stars in common between
  the two datasets. Then, we used the stars in common between
  Pan-STARRS and FORS2 to homogenize the $r$ and $i$ magnitudes of the
  former with the FORS2 magnitudes previously calibrated onto
  HST. Hence, at the end of the procedure, the magnitudes measured in
  the three datasets are all homogeneous and calibrated in the same
  (HST) system. From now on, we use the symbols V$_{\rm 606}$ and
  I$_{\rm 814}$ to indicate both the HST magnitudes, and the
  magnitudes of the other two datasets calibrated onto the F606W and
  the F814W bands, respectively.

\begin{table}
\begin{center}
\begin{tabular}{l|l|l|c|c}
\hline
Instrument &   Program ID & PI  & Filter & $\rm N_{exp} \times \rm T_{exp}$ \\
Survey &   &   & &   \\
\hline
\hline
WFC3 [1]  & GO 11685 & Van Kerkwijk &F606W & 1$\times$392 s + 1$\times$348 s  \\  
 &         &                           & F814W & 1$\times$348 s + 1$\times$261 s   \\ 
\hline
WFC3  [1]  & GO 12517 & Ferraro &F606W & 27$\times$392 s   \\
 &                &              &              F814W & 27$\times$348 s   \\
\hline
WFC3  [1]   & GO 13410 & Pallanca  &F606W & 5$\times$382 s   \\
 &                 &              &              F814W & 5$\times$222 s \\
 &              &              &             F656N & 10$\times$934 s   \\
WFC3  [1]  &        &              &F606W & 5$\times$382 s  \\
&                 &              &              F814W & 5$\times$222 s   \\
 &              &              &              F656N & 10$\times$934 s   \\
WFC3  [1] &         &               &F606W & 5$\times$382 s     \\
&               &              &              F814W & 4$\times$222 s + 1$\times$221 s  \\  
 &              &              &              F656N & 6$\times$934 s + 2$\times$864 s +2$\times$860 s  \\
\hline
WFC3  [1]  & GO/DD 15403 & Pallanca &F606W & 2$\times$382 s   \\
 &                 &              &            F814W & 1$\times$223 s  + 1$\times$222 s  \\
 &              &              &             F656N & 2$\times$969 s + 2$\times$914 s   \\                 
\hline
FORS2  & 077.D-0775(B)& Saviane  &V\_BESS & 1$\times$30 s   \\
 &       &                        &             I\_BESS & 1$\times$30 s \\
\hline
GPC1 [2]  & &   & r &  \\
                     & &   & i & \\
\hline
\end{tabular}
\end{center}
\caption{Summary of the used dataset. [1]=\citet{pallanca+19}, [2]=Pan-STARRS.}
\label{Tab:dataset}
\end{table}

Four catalogs have been obtained from the available datasets.  The
{\it HST-PM} catalog includes all the stars measured in the portion of
the cluster where multi-epoch WFC3 images have been acquired. It has
been corrected for differential reddening \citep[see][]{pallanca+19}
and decontaminated from Galactic field stars via PM analysis.  It
counts 137194 stars.  The {\it HST-noPM} catalog is made of all the
stars detected in the portion of the cluster surveyed by all the
available WFC3 observations. It reports the observed magnitudes (with
no correction for differential reddening) for a total of 174418
objects.  The {\it FORS2} catalog contains 27487 stars measured in a
roughly square region of $200\arcsec$ size around the cluster center.
The {\it Pan-STARRS} catalog lists 40419 stars within a circle of
$700\arcsec$ radius.

Figure \ref{Fig:map} shows the spatial distribution of all the stars
included in each of the four catalogs  (blue dots), with respect
to the cluster center. The corresponding color-magnitude diagrams
(CMDs), which are plotted in Figures \ref{Fig:cmd1} and
\ref{Fig:cmd2}, are deep enough to trace the (cluster and field)
stellar populations down to 3-4 magnitudes below the cluster Red
Clump.  Of course, an increasing population of field stars appears and
becomes dominant with respect to cluster members for increasing
distances from the cluster center, i.e., from the HST, to the FORS2,
to the Pan-STARRS dataset.

\section{Determination of the center of gravity}
\label{center}
As discussed in many previous papers (see, e.g.,
\citealp{montegriffo+95, ferraro+97, ferraro+99a}),  dealing with
resolved stars for the determination of the cluster center avoids
  introducing the bias induced by the possible presence of a few
bright stars, which can generate a SB peak in an
off-set position with respect to the true gravitational center.  Here
we thus took advantage of the {\it HST-PM} catalog, which properly
samples the central region of the cluster, is corrected for the
effects of differential reddening, and has been decontaminated from
field star interlopers (see Figure \ref{Fig:cmd2}).  We used the same
iterative procedure already adopted, e.g., in \citet{lanzoni+07a,
  lanzoni+07b, lanzoni+10, lanzoni+19}, where the gravitational center
($C_{\rm grav}$) is determined by averaging the ($x$, $y$) coordinates
on the plane of the sky of all the stars observed in a selected range
of magnitude and within a circle of radius $r$ centered on a
first-guess value (tipically, the center quoted in the literature).
We always adopt different values for the magnitude-cut and radius $r$,
both to check the occurrence of possible dependencies of the result on
these assumptions, and to estimate the uncertainty on the final
position of $C_{\rm grav}$. As discussed in \citet{miocchi+13}, the
adopted values of $r$ always exceed the core radius ($r_c$) quoted in
the literature, to guarantee that the averaging procedure acts in a
region where the density profile decreases with radius, i.e., is no
more uniform (as it is, instead, in the innermost region). Taking into
account that the literature values of $r_c$ vary from $\sim 5\arcsec$
to $\sim 8\arcsec$ (see Table 2), we adopted $r=15\arcsec$,
$20\arcsec$, $30\arcsec$.  As magnitude-cuts, we used reddening corrected  I$_{\rm 814,0}<18, 18.2,
18.5$, thus selecting approximately equal-mass samples, since the
difference in mass between the stars at the main-sequence turnoff
level and those in evolved evolutionary phases is quite small (within
a few 0.01 $M_\odot$). For every pair of magnitude-cut and $r$ values,
$C_{\rm grav}$ has been determined iteratively starting from the
center quoted in the \citet[][2010 edition]{harris} catalog and
assuming that convergence is reached when ten consecutive iterations
yield values of the cluster center that differ by less than
$0.01\arcsec$ among them. As gravitational center of NGC 6440, we
finally adopted the average of the values of $C_{\rm grav}$ obtained
from this procedure, namely: $\alpha = 17^{\rm h} 48^{\rm m}
52.84^{\rm s}$ and $\delta = -20^{\circ} 21\arcmin 37.5\arcsec $, 
  with an uncertainty of $\sim 0.3\arcsec$.  This is $\sim 2\arcsec$
east and $\sim 0.6\arcsec$ south from the center quoted in the
\citet{harris} catalog.  Such a difference can have a non negligible
impact on the derived shape of the star density profile and, more in
general, on the study of the radial behavior of any stellar population
within the cluster potential well.

\section{Star count density profile}
\label{dens}
\begin{figure*} 
\begin{center}
\includegraphics[width=150mm,angle=0]{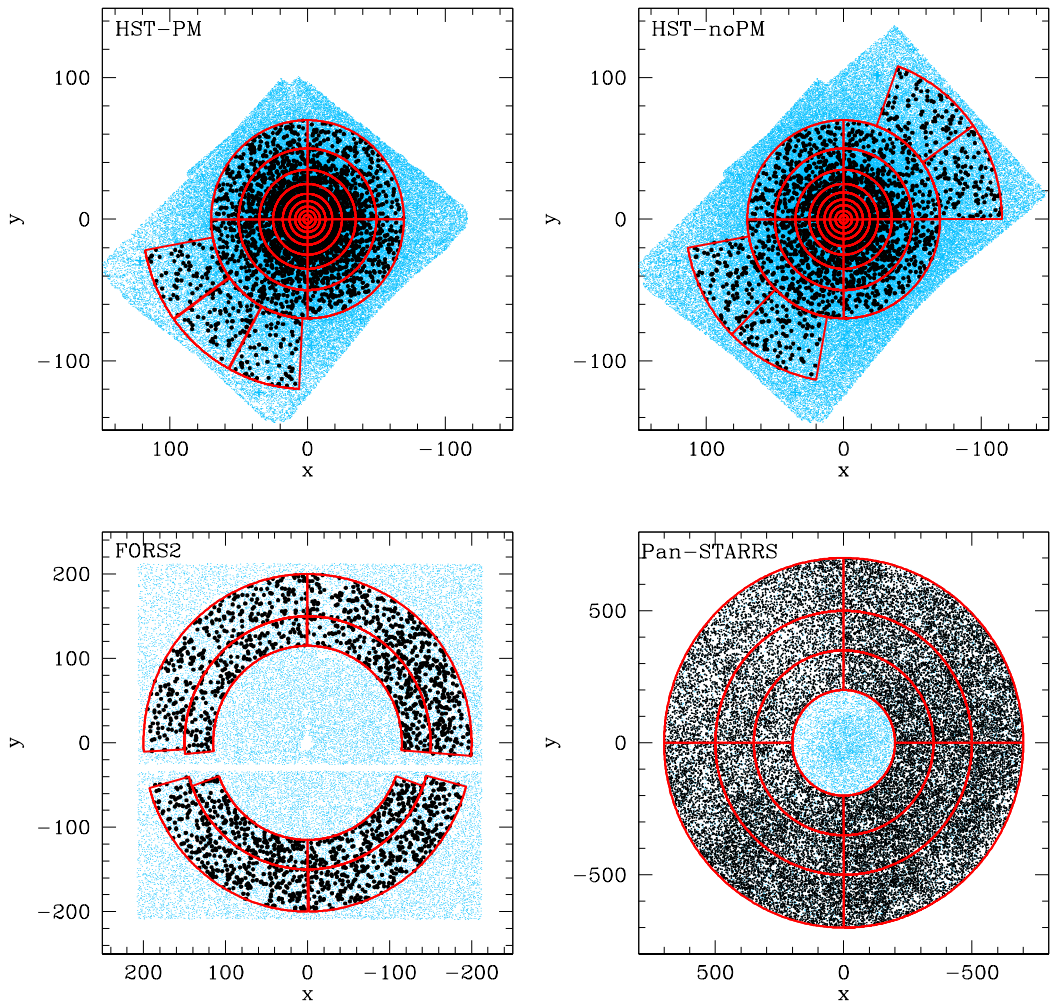}
\caption{Spatial distribution of  all the observed stars in each
    catalog (blue dots), with respect to the newly determined cluster
  center (see Table \ref{tab_params}).  The black dots correspond
    to the objects used to determined the cluster density profile:
    stars with I$_{\rm 814}<19$ beloging to the HST-PM catalog (used
    to build the density profile shown in red in Figure
    \ref{Fig:prof}) and stars with I$_{\rm 814}<18.5$ in the other
    three datasets (used for the density profile shown in black in
    Figure \ref{Fig:prof}). The red lines delineate the annuli and
    sub-sectors used to construct the density profile.}
\label{Fig:map}
\end{center}
\end{figure*}

\begin{figure*} 
\begin{center}
\includegraphics[width=150mm,angle=0]{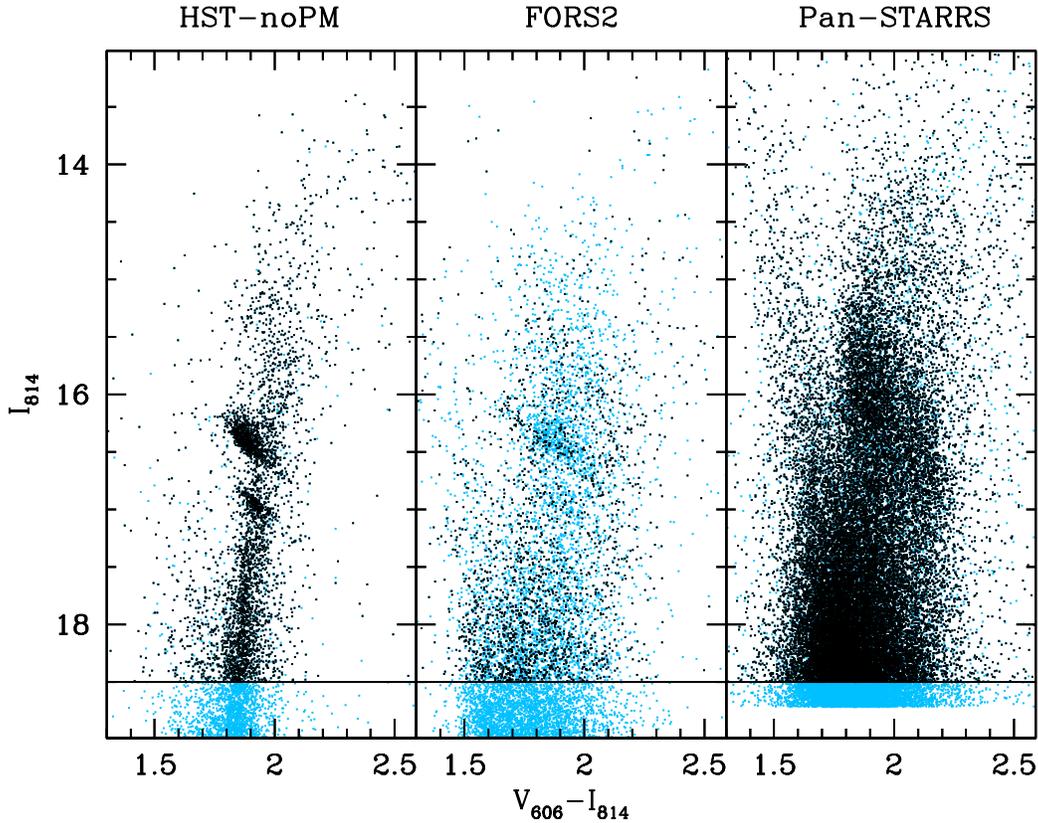}
\caption{CMDs of NGC 6440 obtained from the {\it HST-noPM}, {\it
    FORS2} and {\it Pan-STARRS} catalogs discussed in the text (left,
  central, and right panel, respectively). The entire samples of
  surveyed stars are plotted in  light blue, while the black dots highlight
  the stars used for the construction of the density profile. }
\label{Fig:cmd1}
\end{center}
\end{figure*}

\begin{figure*} 
\begin{center}
\includegraphics[width=150mm,angle=0]{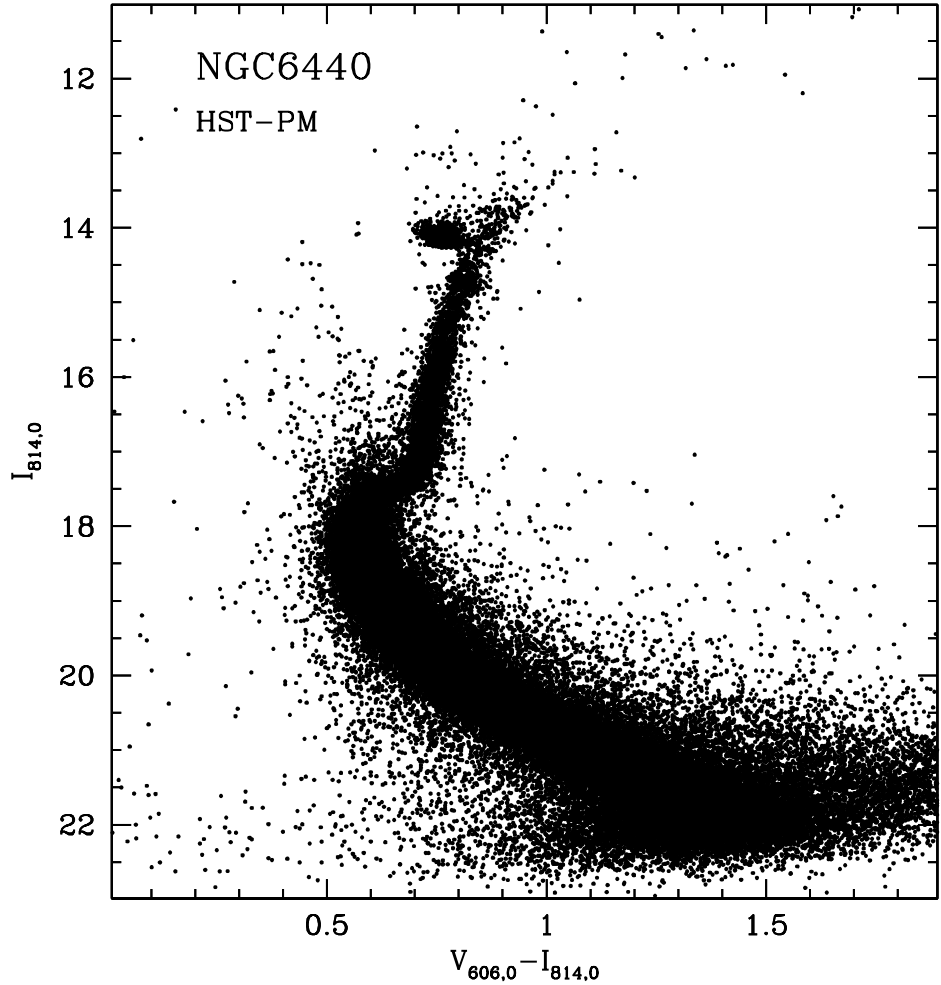}
\caption{CMD of NGC 6440 from the {\it HST-PM} catalog. Only
  high-quality stars are plotted. Non-member stars (as determined from
  the proper motion analysis) have been removed. The magnitudes are
  corrected for the effect of reddening \citep{pallanca+19}.  }
\label{Fig:cmd2}
\end{center}
\end{figure*}

In order to build the projected star density profile, $\Sigma_*(r)$,
along the entire cluster radial extension, we combined the available
photometric data-sets as follows: the {\it HST-noPM catalog} covers
the innermost cluster regions ($\le115\arcsec$), the {\it FORS2
  catalog}, where the center remains unresolved because of stellar
crowding, is used to sample the intermediate regions
($115\arcsec<r\le200\arcsec$), and the {\it Pan-STARRS catalog} refers
to the outermost cluster regions ($200\arcsec<r\le700\arcsec$).  We
considered only stars brighter than I$_{\rm 814}<18.5$ (i.e., $\sim 2$ mag above
the main-sequence turnoff;  black dots in the top-right and bottom
  panels of Figure \ref{Fig:map} and in Figure \ref{Fig:cmd1}),
because this limit ensures comparable levels of (high) photometric
completeness, in combination with high enough statistics (thousands of
stars) in all the catalogs.  Following the standard procedure already
adopted in several previous works (see \citealp{lanzoni+19} and
references therein), we divided each photometric sample in several
concentric annuli centered on $C_{\rm grav}$ (see Figure
\ref{Fig:map}), and split each annulus into an adequate number of
sub-sectors (typically four). The number of stars lying in each
sub-sector was counted, and the star surface density was obtained by
dividing these values by the corresponding sub-sector area. The
stellar density in each annulus was then obtained as the average of
the sub-sector densities, and the standard deviation was adopted as
the uncertainty.
   
\begin{figure*} 
\begin{center}
\includegraphics[width=150mm]{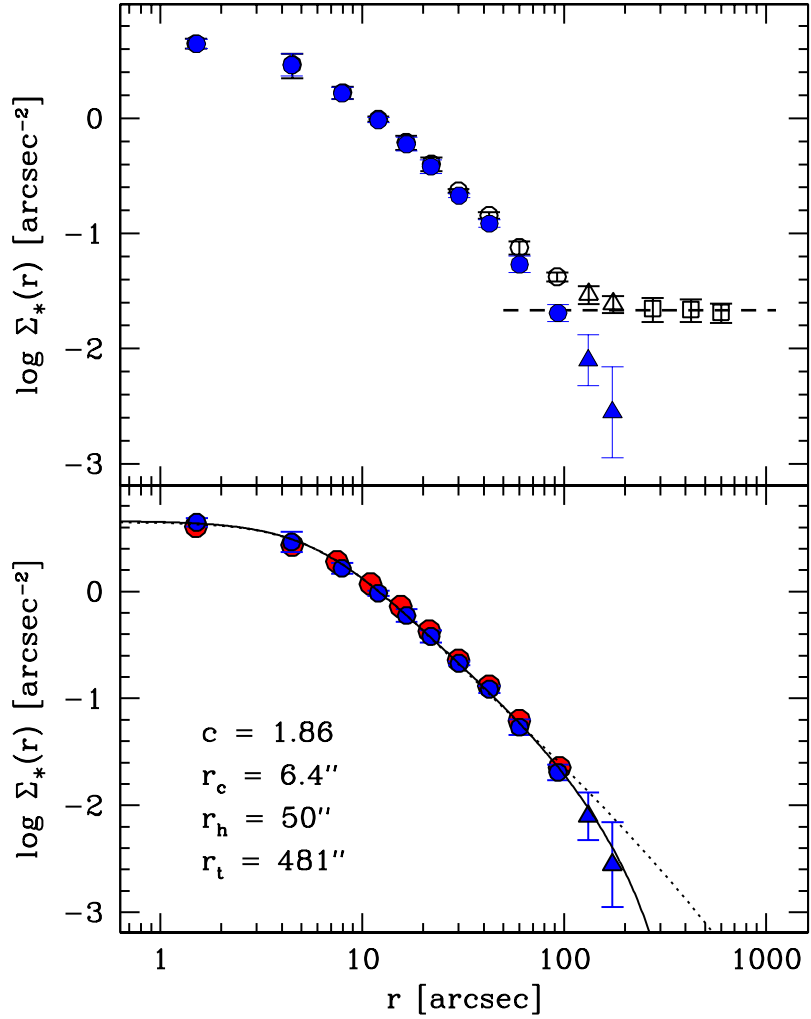}
  \caption{{\it Top:} Observed star density profile of NGC 6440
    obtained from resolved star counts by combining three
      different catalogs: {\it HST-noPM} (empty circles), {\it FORS2}
    (empty triangles) and {\it Pan-STARRS} (empty squares). The filled
    blue symbols correspond to the cluster density profile obtained
    after subtraction of the Galaxy field contribution (dashed
    line). {\it Bottom:} Cluster density profile shown in the top
    panel (blue symbols) compared to that obtained from the {\it
      HST-PM} catalog (red circles). As can be seen the agreement is
    very good, thus guaranteeing the reliability of the applied
    background subtraction (see text).  The black line shows the
    best-fit King model profile. The corresponding values of the
    concentration parameter ($c$) and a few characteristic
    scale-lengths (in arcseconds) are also labelled. The dotted
      line shows the best-fit Wilson model.}
\label{Fig:prof}
\end{center}
\end{figure*}

The observed stellar density profile is shown in Figure \ref{Fig:prof}
(upper panel), where different symbols refer to different catalogs
(empty circles for HST, triangles for FORS2 and squares for
Pan-STARRS) and the radius associated with each annulus is the
midpoint of the radial bin. As can be seen, the contribution of the
Galactic field starts to be evident for distances from the cluster
center larger than $r>100\arcsec$ (i.e. in the FORS2 data-set) and
becomes dominant for $r>200\arcsec$ (Pan-STARRS catalog). As expected,
the spatial distribution of field stars is approximately uniform on
the considered radial scale, thus producing a well defined plateau in
the outermost portion of the density profile. Hence,  the  level
of Galactic field contamination has been estimated by averaging the
data-points aligned in the plateau (see the dashed line in Figure
\ref{Fig:prof}) and the (decontaminated) cluster profile, obtained
after subtraction of the Galaxy background level, is finally shown in
Figure \ref{Fig:prof} (top panel; filled symbols). As apparent,
after the field subtraction, the profile remains almost unchanged at
small radii, which are in fact dominated by the cluster population,
while it significantly decreases in the most external regions, where
it turns out to be significantly below the Galactic background. This
clearly indicates that an accurate measure of the field level is
crucial for the reliable determination of the outermost portion of the
density profile.

The background subtraction has a well-perceivable effect also in the
region sampled by HST data. We thus took advantage of the {\it HST-PM}
catalog, which is already cleaned from Galactic field interlopers, to
double check the reliability of the adopted decontamination procedure.
Thanks to the high level of completeness of the HST observations, we
constructed the cluster density profile by using stars down to the
sub-giant branch (i.e.,  0.5 magnitudes deeper than the sample
used in the previous procedure;  black dots in the top-left panel
  of Figure \ref{Fig:map}), thus benefitting from a much larger
statistics. Figure \ref{Fig:prof} (bottom panel) shows the density
profile obtained from the {\it HST-PM} catalog (in red), vertically
shifted to match the one obtained with the procedure described above
(in blue): as can be seen, the two profiles are essentially identical
in the common region, thus confirming the solidity of the adopted
field decontamination approach.

In order to derive the physical parameters of the program cluster, we
fit the observed star density profile with the family of \cite{king66}
models in the isotropic, spherical and single-mass
approximation.
They constitute a \emph{single}-parameter family, since their shape is
uniquely determined by the dimensionless parameter $W_0$, which is
proportional to the gravitational potential at the center of the
system, or, alternatively, to the ``concentration parameter'' $c$,
defined as $c\equiv\log(r_t/r_0)$, where $r_t $ and $r_0$ are the
tidal and the King radii of the cluster, respectively.

The best-fit King model has been determined by exploring a grid
  of $W_0$ values varying between 0.4 and 12 in steps of 0.05, and
  selecting the solution that minimizes the $\chi^2$ residuals between
  the observed and the theoretical density profiles (see
  \citealp{miocchi+13} and \citealp{lanzoni+19} for a detailed
  description of the adopted procedure and the method used to estimate
  the uncertainties). The resulting values of $W_0$, concentration
  parameter, core, half-mass and tidal radii are: $W_0 = 8.10$, $c =
  1.86$, $r_c=6.4\arcsec$, $r_h=50.2\arcsec$, $r_t=481.4\arcsec$,
  respectively, with the uncertainties quoted in Table
  \ref{tab_params}. The effective radius, defined as the radial
distance including half the total number counts in projection (and
corresponding to the projected half-light radius if SB, instead of
resolved star density, is considered) is $r_e = 36.8\arcsec$.  The
comparison with previous determinations in the literature shows
significant discrepancies for all the parameters.  In particular, the
Harris catalog reports the values estimated by \citet{mclaughlin+05},
who found $c=1.62$, $r_c=8.1\arcsec$, $r_h=44.8\arcsec$, $r_{\rm
  eff}=28.8\arcsec$, and $r_t=354.9\arcsec$ (the values originally
quoted in pc have been converted into arcseconds by using the cluster
distance provided in that paper: $d=8.4$ kpc). Hence, we find that NGC
6440 is more centrally concentrated than previously thought, with a
smaller core radius and larger truncation radius, translating in a
larger concentration parameter.  The comparison with the values quoted
by \citet{baumgardt+18} is less straightforward because N-body
simulations, instead of King models, are used there to fit the
observation, and no uncertainties are provided. However, the provided
values of core, half-mass and effective radii are consistent with ours
within 10-20\%. We also stress that SB (instead of number count)
profiles and an offset position of the cluster centre are used in
those studies, thus likely accounting for the different results. The
density profile of some GCs is found to be best reproduced by
\citet{wilson75}, instead of \citet{king66}, models \citep[see,
  e.g.,][]{mclaughlin+05, miocchi+13}. We thus compared the
observations also with the \citet{wilson75} model family, finding the
best solution for $c=3.30$, $r_c=6.6\arcsec$ and $r_t\sim 230\arcmin$
(see the dotted line in the bottom panel Figure
\ref{Fig:prof}). Although the core radius is very similar to that
obtained from the King fit, the Wilson model (that, by construction,
provides a smoother cutoff at the limiting radius) severely
overestimate the observed stellar density in the external portion of
the profile. This is in agreement with the fact that NGC 6440 is
orbiting the Bulge of our galaxy, where tidal truncation is expected
to be more relevant than for faraway halo GCs.

\section{The distance modulus and the age of NGC 6440}
\label{DMage}
The distance modulus and the age of resolved stellar populations (as
Galactic GCs) can be estimated through the comparison between the
observed CMD and theoretical stellar isochrones, the main obstacle
being the well known degeneracy of these parameters with the
metallicity and the reddening.

In the case of NGC 6440, the metallicity is relatively well known
since the first low-resolution spectroscopic measures (see
\citealt{arma+88, origlia+97, frogel+01}), indicating an overall iron
abundance of the order of 1/3 - half solar. More recent
high-resolution spectroscopy of small samples of giants measured in
the IR \citep{origlia+08met} and in the optical band \citep{munoz+17}
confirmed a considerable iron content ([Fe/H]$=-0.5, -0.6$) with some
$\alpha-$enhancement ($[\alpha/Fe]=+0.3$), corresponding to a global
metallicity\footnote{ The global metallicity has
been calculated trough the relation reported by \citet{ferraro+99b}
and assuming $[Fe/H]=-0.56$ \citep{origlia+08met}} [M/H]$\sim -0.4$.

The accurate estimate of the reddening, instead, is complicated by the
fact that NGC 6440 is located close to the Galactic plane and toward
the bulge, where the extinction law likely deviates from the canonical
and commonly assumed behavior.
 An extensive discussion about extinction and reddening is
  presented in \citet{mccall04}.  Particular attention, however, has
  to be given to the extinction toward the inner Galaxy, where the
  R$_V$ value is not constant and can significantly vary along
  different directions \citep[e.g.,][and references
    therein]{popowski00, udalski03,nataf13,alonsoGarcia17}.  Indeed,
as discussed in \citet{nataf13, casagrande14}, and recently confirmed
by \citet{ferraro+21} and C. Pallanca et al. (2021, in preparation),
the region toward the Galactic center seems to be better described by
an extinction law with a significantly smaller value of $R_V$ (even
down to $R_V=2.5$).   Conversely some other authors found larger
  $R_V$ values ($R_V=3.2$) to be more appropriate
  \citep[e.g.,][]{bica16,kerber19}.  Hence, as already discussed by
  \citet{udalski03}, the proper dereddening of a particular field in
  the Galactic bulge might be difficult without prior determination of
  $R_V$ along its line of sight.

The best way to constrain $R_V$  in a specific direction is by
simultaneously investigating the IR and the optical CMDs, which are,
respectively, weakly and strongly sensitive to the true extinction
law. To this purpose we used a combination of optical and IR catalogs
of NGC 6440.

\begin{figure*} 
\begin{center}
\includegraphics[width=150mm]{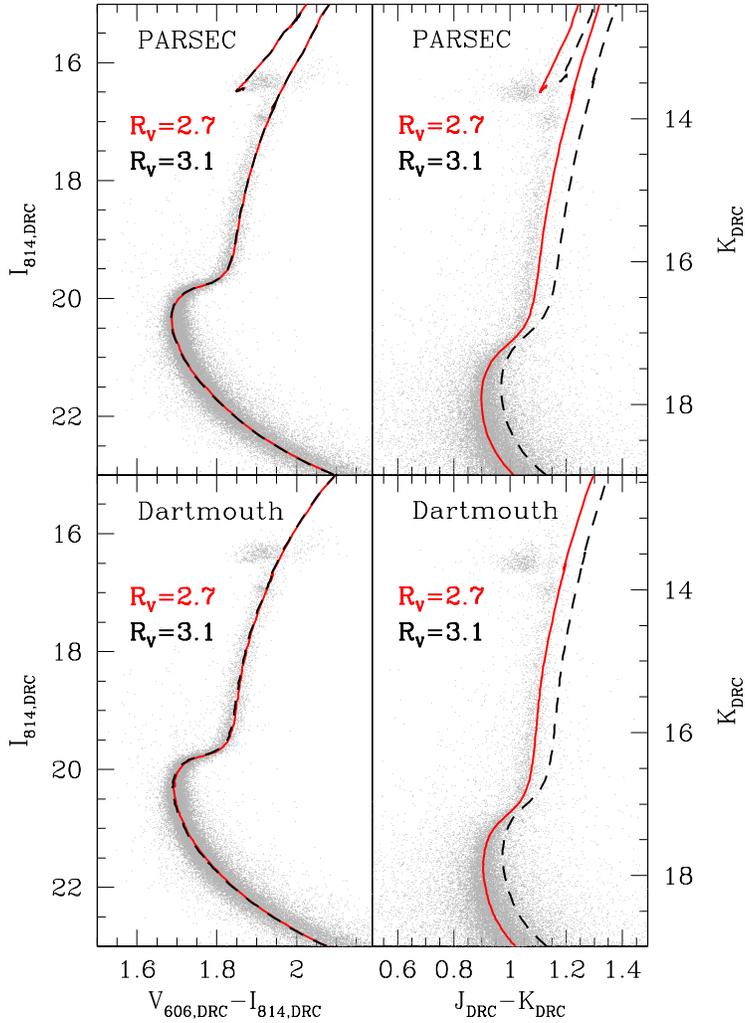}
\caption{Comparison between the differential-reddening corrected CMD
  of NGC 6440 (grey dots) and theoretical isochrones computed for
  $R_V=3.1$ (black dashed lines) and for $R_V=2.7$ (red lines). The
  left and right panels show, respectively, the optical and the IR
  CMDs. The top panels refer to PARSEC isochrones, while the bottom
  panels show Dartmouth models. No solution able to properly fit the
  optical and the IR CMDs simultaneously is found for $R_V=3.1$, while
  if $R_V=2.7$ is assumed, both models well reproduce the observations
  for an age of 13 Gyr, a distance modulus of 14.60 and
  $E(B-V)=1.26$-1.28.   The fact that the RGB looks steeper than
    the model in the IR CMDs may be due to nonlinearity effects of the
    GeMS/GSAOI photometry \citep[see][]{saracino+16}.  }
\label{Fig:rvcmd}
\end{center}
\end{figure*}

 The adopted HST optical catalog was presented in
  \citet{pallanca+19}, while the IR one is based on deep $J$ and $K_s$
  observations obtained with GeMS/GSAOI (S. Saracino et al., 2021, in
  preparation). For a proper comparison with stellar isochrones, we
  first corrected the CMDs obtained from these catalogs for the effect
  of differential reddening, which broadens and distorts the
  evolutionary sequences. To this end, we applied the procedure fully
  described in \citet{pallanca+19} to the HST dataset. Briefly, we
  determined the reference mean ridge line of NGC 6440 using a sample
  of well-measured stars. Then, for every star in the HST catalog we
  selected a sample of close sources, thus defining a
  ``local-CDM''. Finally, we estimated the value $\delta$E(B-V)
  necessary to superpose the reference mean ridge line onto the
  local-CMD and assigned this value to the corresponding investigated
  star. By construction, the $\delta$E(B-V) values thus obtained
  express the differential component of the reddening within the
  sampled FOV and can be positive or negative. This quantity, multiplied by the coefficient appropriate for the 
  considered filter, is added to the observed magnitudes to get differential 
  reddening corrected (DRC)  magnitudes: I$_{\rm 814,DRC}$, V$_{\rm
    606,DRC}$ (see Figure \ref{Fig:rvcmd}). Finally, for all the stars in
  common with the GeMS/GSAOI sample, the estimated values of
  $\delta$E(B-V) have been used to correct also the IR magnitudes and
  build the corresponding differential reddening corrected CMD:
  K$_{\rm DRC}$, (J-K)$_{\rm DRC}$ (Figure \ref{Fig:rvcmd}).

To estimate the distance modulus and the age of NGC 6440,  while
  constraing $R_V$, we then compared the differential reddening
corrected CMDs with a set of PARSEC \citep{bressan+12} and
Dartmouth \citep{dotter+08} isochrones of different ages
computed for [M/H]$=-0.4$
in the four photometric bands of interest,
namely, the WFC3  V$_{\rm 606}$ ad I$_{\rm 814}$ filters, and the 2MASS $J$ and $Ks$
filters.  To place the isochrones in the differential reddening
corrected CMDs, we determined the values of the (temperature- and
gravity-dependent) extinction coefficients in the four bands (namely,
$R_{\rm V_{606},i}$, $R_{\rm I_{814},i}$, $R_{J,i}$ and $R_{Ks,i}$) for each $i^{th}$
combination of effective temperature and surface gravity along every
isochrone \citep{casagrande14}. This has been done by interpolating
the values calculated for the MARCS grid \citep{marcs} under the
assumption of the cluster metallicity and the \citet{cardelli}
  extinction law with $R_V=2.5$ and $R_V=3.1$.
A linear interpolation between these two limits then allowed us to
determine the temperature- and gravity-dependent coefficients for
intermediate values of $R_V$ (between 2.5 and 3.1, stepped by 0.1).

\begin{figure*} 
\begin{center}
\includegraphics[width=150mm]{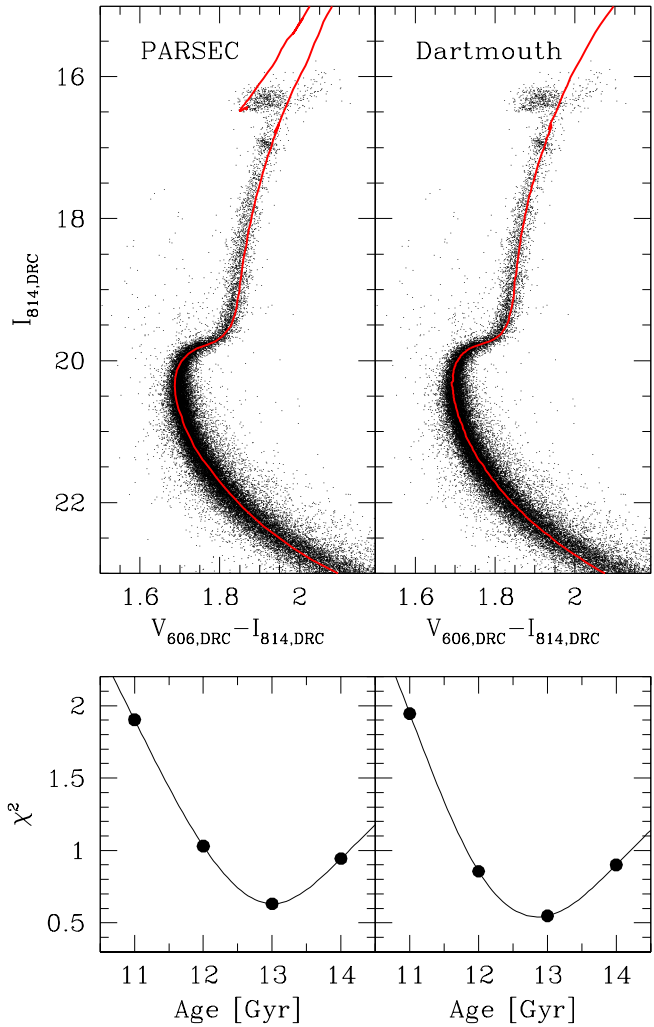}
\caption{{\it Top panels:} differential-reddening corrected CMD of NGC
  6440 with the best-fit isochrone from the PARSEC (left) and the
  Dartmouth (right) models superimposed as red lines (the same as
    in Figure \ref{Fig:rvcmd}). {\it Bottom panels:} value of the
  $\chi^2$ parameter (see text) obtained from the fit of the SGB/MS-TO
  region through PARSEC (left) and Dartmouth (right) isochrones of 11,
  12, 13, and 14 Gyr, as a function of the model age. The minimum of
  the $\chi^2$ parameter is found for an age of 13 Gyr.}
\label{Fig:eta}
\end{center}
\end{figure*}

We found that, under the assumption of $R_V=3.1$, no combination of
parameters is able to reproduce the optical and the IR CMDs
simultaneously. In particular, a combination that well fits the
optical CMD corresponds to an excessively bright and red isochrone in
the IR, for both the considered models (see the black dashed lines in
Figure \ref{Fig:rvcmd}). The problem becomes progressively milder for
decreasing values of $R_V$, and the best solution is found for
$R_V=2.7$. Interestingly, similarly small values are needed to
properly reproduce the observations of other GCs in the Galactic bulge
\citep[][see also Pallanca et al. 2021, in preparation]{ferraro+21}.
The red lines shown in Figure \ref{Fig:rvcmd} correspond to PARSEC and
Dartmouth isochrones (upper and lower panels, respectively) computed
under the assumption of $R_V=2.7$, for an age of 13 Gyr, a distance
modulus $\mu_0=14.60$ and an absolute color excess that is just
slightly different between the two models: $E(B-V)=1.28$ and 1.26,
respectively. In particular, this combination of parameters not only
well reproduces the horizontal branch magnitude level (see the PARSEC
isochrones in the figure), but also best-fits the SGB/MS-TO region,
which is the most sensitive to age variations.  The best-fit solution
has been evaluated through a $\chi^2$ analysis, by determining, for
each SGB/MS-TO star, the difference between its observed color and the
color at the same magnitude level along the isochrones of 11, 12, 13
and 14 Gyr. As already done in previous papers (\citealp{ferraro+21};
see also \citealp{saracino+16}) the $\chi^2$ parameter has been
computed as the ratio between the square of this difference and the
color along the isochrone, summed over all the selected stars. The
best-fit model to the optical CMD and the $\chi^2$ values as a
function of the investigated ages are plotted, respectively, in the
top and bottom panels of Figure \ref{Fig:eta}.  Taking into account
the various uncertainties and degeneracies entering the fitting
procedures, conservative estimates of the errors on the distance
modulus and age are 0.1 mag and 1.5 Gyr, respectively.

 Several previous works in the literature have been devoted to the
  determination of these parameters for NGC 6440.  The study by
  \citet{ortolani94} is based on optical photometry, while the others
  have been performed in the IR, and in  all cases the standard reddening
  law has been assumed. The only exception is the investigation of
  \citet{kuchinskiFrogel}, who combined IR data with V band photometry
  and suggested that the anomalous colors observed for this cluster
  might require a non-standard reddening law. This is in agreement
  with our finding, although a detailed comparison between the
  proposed reddening laws in not obvious. Given the different $R_V$
  adoption, the comparison among various reddening determinations in
  the literature has to be done in terms of the extinction coefficient
  $A_V$, instead of the color excess E(B-V), which is linked to the
  former by the following relation: $A_V = R_V \times$ E(B-V).
  \citet{minniti95} quote E(J-K)=0.57 and comment that this value is
  in good agreement with that of \citet{ortolani94}, who found
  E(B-V)=1 and adopt $R_V=3$, corresponding to $A_V =
  3$. \citet{valenti+04} quote E(B-V)=1.15 and adopt $R_V=3.1$, thus
  providing $A_V=3.56$. From the Harris compilation, $A_V=3.32$ is
  obtained for the standard value of $R_V$. The value estimated in the
  present study ($A_V=2.7 \times1.27=3.43$) therefore is within the
  range spanned by the results of previous works, which however do not
  correct for differential reddening, nor take into account the
  optical and IR CMDs simultaneously. Also the distance modulus here
  determined is in reasonable agreement with previous determinations
  and included between them: $\mu_0= 14.64$ and 14.58 in
  \citet{ortolani94} and \citet{valenti+04}, respectively. Finally,
  an  age of $11^{+3}_{-2}$ Gyr was estimated by \citet{origlia+08} from
  pioneering adaptive optics photometry, and 13 Gyr is the value
  adopted by \citet{munoz+17} to fit the observed CMD with theoretical
  isochrones. These are both consistent with our determination of
  $13\pm 1.5$ Gyr.

\section{The RGB-bump}
\label{bump}
\begin{figure*} 
\begin{center}
\includegraphics[width=150mm]{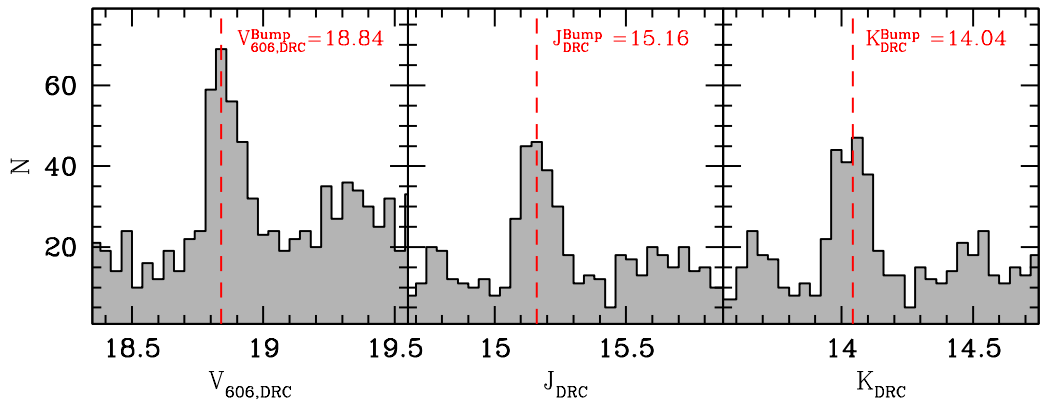}
\caption{Differential luminosity function of RGB stars classified as
  cluster members and photometrically well measured. The detected
  peaks (marked by the dashed red lines) correspond to the RGB-bump
  magnitude in the three photometric bands (see labels). }
\label{Fig:bump1}
\end{center}
\end{figure*}

\begin{figure*} 
\begin{center}
\includegraphics[width=150mm]{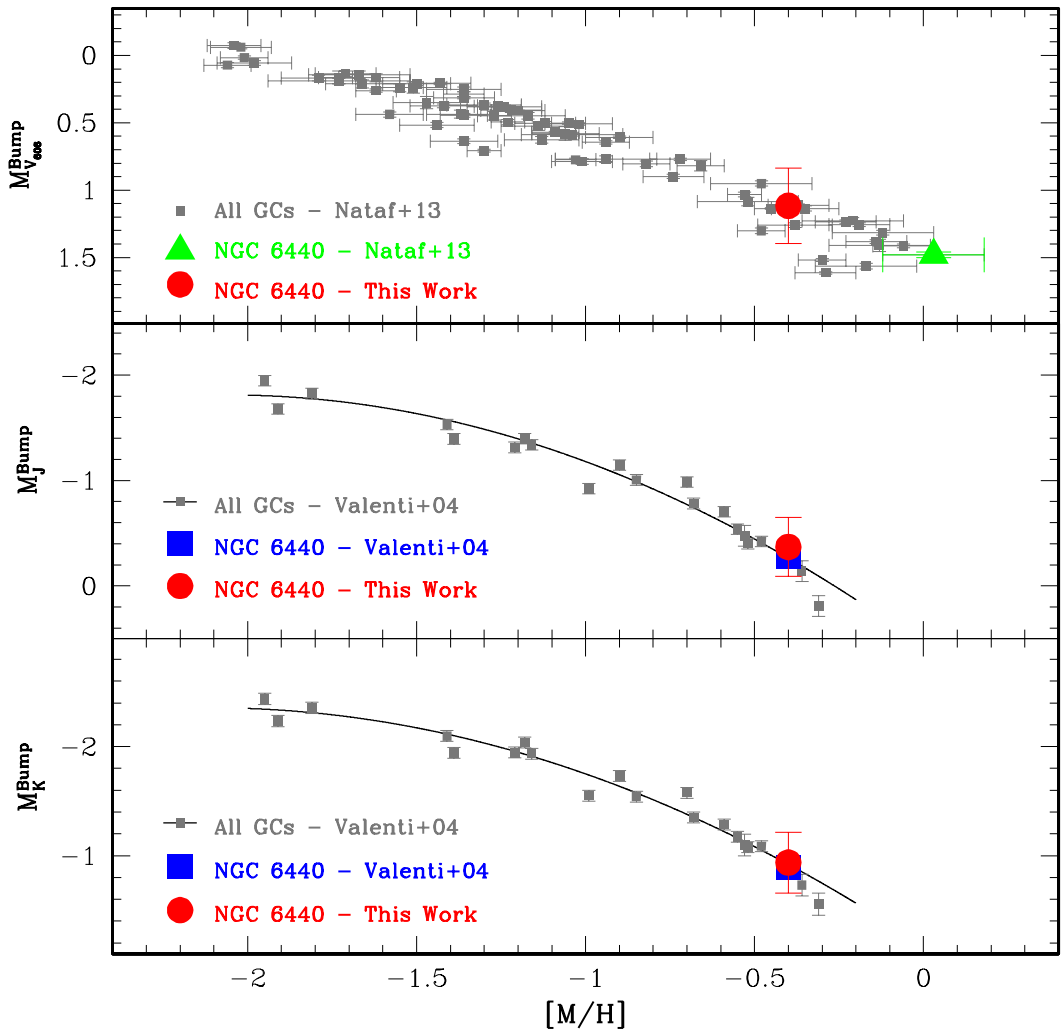}
\caption{Absolute magnitude of the RGB-bump in the V$_{\rm 606}$, J and K bands
  (from top to bottom) as a function of the GC global metallicity
  [M/H]. The gray symbols are from the literature (see labels). The
  blue squares and the green triangle mark the location of NGC 6440
  according to literature \citep{nataf+13bump,valenti+04bump} while
  the red circle mark the values determined in this
  work. }\label{Fig:bump2}
\end{center}
\end{figure*}

The high-quality of the CMDs presented in this paper allows us to
easily identify a well known evolutionary feature along the RGB: the
so-called RGB bump.  This feature appears in the CMD as a well
defined clump of stars along the RGB. This evolutionary feature flags
the moment when the H-burning shell reaches the H discontinuity left
by the inner penetration of the convective envelope (see the seminal
works by \citealt{fusipecci+90}, \citealt{ferraro+91, ferraro+92a,
  ferraro+92b} and \citealt{ferraro+99b,ferraro+00}; see also the
compilation by \citealt{zoccali+99}, \citealt{riello+03} and
\citealt{valenti+04bump}, and more recently by
\citealt{nataf+13bump}).  Figure \ref{Fig:bump1} shows the
differential luminosity function of the bright RGB stars in the
differentially-corrected V$_{\rm 606}$, $J$, $K$ bands. The well-defined peaks
 at  V$_{\rm 606,DRC}= 18.84\pm0.05$, $J_{DRC}= 15.16\pm0.05$, $K_{DRC}=
14.04\pm0.05$ correspond to the RGB bump. Adopting the extinction and the
distance reported in Table \ref{tab_params}, we transformed the above
values into absolute magnitudes  obtaining $M^{Bump}_{\rm V_{606}} =1.12 \pm0.12$,
$M_{J}^{Bump}=-0.37\pm0.12 $ and $M_{K}^{Bump}=-0.94\pm0.12 $.  In
Figure \ref{Fig:bump2} we show the comparison among these measures and
previous determinations in the literature. In particular in the bottom
and central panels of Figure \ref{Fig:bump2} we show the nice
agreement of the RGB bump magnitude in the IR bands with the estimates and
the relations quoted in \citet{valenti+04bump}.  The situation appears
to be much more complex in the optical V$_{\rm 606}$ band, since the magnitude
level found in the present study appears significantly brighter than
that obtained by \citet{nataf+13bump}, who quote
 $M_{\rm V_{606}}^{Bump}=1.48$. While the adopted extinction law and distance
modulus are just slightly different between the two studies, most of
the discrepancy is due to the observed RGB bump magnitude:
V$_{606}=19.431\pm0.021$ in \citet{nataf+13bump}, V$_{606}=18.9$ in our
study. In addition, a metallicity as high as [M/H]=+0.03 has been
assumed in that work.

Nevertheless, we note that the new determination of the RGB bump combined
with the adopted metallicity well follows the trend of the GC
distribution reported by \citet{nataf+13bump}.

\section{SUMMARY AND CONCLUSIONS}
\label{conclu}
This work
 provides updated values for the structural parameters  and
  age of NGC 6440, a GC in the direction of the Galactic bulge that
is relatively poorly investigated because of the strong and
differential extinction along its line of sight.  To our knowledge,
these are the first determinations of the center and density profile
of NGC 6440 based on resolved star counts, which are free from
 biases induced by the possible presence of few bright
objects. The gravitational center of the cluster has been determined
from the observed positions of PM-selected member stars, and it turns
out to be significantly off (by $\sim 2\arcsec$ in right ascension)
with respect to the value quoted in the literature \citep{harris},
reporting the SB peak estimated by \citet{picard95}. By making use of
a suitable combination of high-resolution (HST) photometry and
wide-field data (FORS2 observations and a Pan-STARRS catalogue), we
then built the most radially extended surface density profile so far
from resolved star counts. With respect to the work of \citet[][see
  also the Harris catalog]{mclaughlin+05}, which is based on the
previous estimate of the cluster center and uses the SB distribution
instead of number counts, the best-fit King model to the projected
density profile derived in this work reveals that NGC 6440 has a
significantly larger concentration parameter, a smaller core radius,
and a larger overall extension (truncation radius) on the plane of the
sky. The updated values of the cluster center and structural
parameters are listed in Table \ref{tab_params}.

By taking advantage of the PM selection, we built a sample of cluster
member stars with both optical (HST) and IR (GeMS/GSAOI) magnitudes,
 that we properly corrected for the effect of differential
  reddening. We then used stellar isochrones, from both the PARSEC
\citep{bressan+12} and the Dartmouth \citep{dotter+08} models, to
simultaneously reproduce the optical and the IR CMDs  (which are
  strongly and weakly dependent on reddening, respectively). 
Adopting
  extinction coefficients that depend on the stellar surface
  temperature and gravity, we explored extinction laws with $R_V$
  ranging from 2.5 to 3.1, and we found that $R_V=2.7$ is required to
  fit both to the optical and the IR CMDs.  This confirms 
  \citep[see also][]{popowski00,nataf13, casagrande14} that the extinction law in Galactic regions close to the
  plane and in the direction of the bulge requires an $R_V$ value
  significantly different from the ``canonical'' 3.1. The best-fit to
  the CMDs (in particular to the horizontal branch level, and the
SGB/MS-TO region that strongly depends on the age of the stellar
population) provided us also with the cluster age ($t= 13 \pm 1.5$
Gyr), distance modulus ($\mu_0=14.60 \pm 0.1$, corresponding to a
distance of 8.3 kpc, with a conservative uncertainty of $\sim 0.4$
kpc),  and absolute color excess, $E(B-V)=1.26-1.28$, which
  corresponds to a V-band extiction coefficient $A_V=3.34$. These
  values are all within the ranges spanned by previous determinations
  in the literature (see Section \ref{DMage}).
  In particular, the age
  estimate here obtained for NGC 6440 is the most accurate so far
  (although the uncertainty is still quite large: 1.5 Gyr).
 Figure 9 shows the age-metallicity distribution for the bulge GCs with available age
  estimate, where NGC 6440 is marked as a large red square. 
  The data for the other clusters are mainly from 
    \citet[][see their Figure 16]{saracino+19}  and \citet[][see their Figure 12]{oliveira+20}  with the   
    addition of the recent age determination of NGC 6256 \citep{cadelano+20}. 
    We also mark the age-metallicity of the oldest
stellar population in the two Bulge Fossil Fragments (BFF; namely Terzan5 and Liller1) so far
discovered into the bulge \citep{ferraro+09b,ferraro+16b,ferraro+21}. 
The BFFs  \citep{ferraro+09b, ferraro+16b, origlia+11, massari+14} are systems that, in spite of their appearance as genuine GCs, host multi-age stellar populations and could be the remnants of massive clumps that contributed to form the bulge at the epoch of the Galaxy assembly. As apparent, these systems all have old ages, well consistent with those of the majority of bulge GCs and Galactic field stars observed in different directions toward the bulge \citep[e.g.,][]{zoccali+03,
    clarckson+11, valenti+13}. The  weighted mean age  of the entire sample (now including a total  of 18 GCs and 2 BFFs) is $12.7 \pm 0.2$ Gyr, which is $\sim 0.5$ Gyr older than the value quoted in \citet{saracino+19} on the basis of a sub-sample of 14 objects.

The superb quality of the obtained CMDs allows an accurate determination of the RGB-bump. This value, 
combined with the spectroscopic estimate of the cluster metallicity, makes NGC 6440 to perfectly fit into the 
bump-metallicity relation defined by Galactic GCs.
  
The new determinations of the cluster structural parameters, distance
and reddening allow us to also update the value of relaxation time of
NGC 6440, which characterizes the dynamical evolutionary stage of the
system. It quantifies the timescale needed by two-body interactions
(causing kinetic-energy exchanges among stars) to bring the cluster
toward a thermodynamically relaxed state. This quantity has been used
to validate the so-called $A^+$ parameter, quantifying the level of
central segregation of blue straggler stars within a GC
\citep{alessandrini+16}, as a powerful empirical diagnostic of the
dynamical age of the host system \citep[e.g.,][]{ferraro+18a,
  ferraro+19, lanzoni+16, ferraro+20}. Primarily depending on the
local density, the value of the relaxation time changes with the
radial distance from the cluster center.  To estimate the central
relaxation time ($t_{rc}$) we used equation (10) of \citet{djo93}. For
the half-mass relaxation time ($t_{rh}$) we followed equation (8-72)
of \citet{BT87}. The latter parameter can be estimated also from
eq. (11) of \citet{djo93} once the first coefficient ($8.993\times
10^5$ in the equation) is substituted by its proper value ($2.055
\times 10^6$; see \citealp{BT87, mclaughlin+05}). We also emphasize
that the projected half-light radius $r_e$ (instead of the
three-dimensional half-mass radius $r_h$) is often used in this
estimate (see, e.g., \citealp{harris, mclaughlin+05}), under the
implicit assumption that these radial scales are equal. However,
depending on the value of the concentration paramenter, the ratio
between the effective and the half-mass radii varies between 0.73 and
0.76. As a consequence, since $t_{rh}$ scales with half-mass radius at
the power of 3/2, the relaxation time obtained by adopting $r_e$
($t_{re}$) is $\sim 35\%$ shorter than that calculated by using
$r_h$. Assuming the absolute $V-$band magnitude and central SB quoted
for NGC 6440 in the \citet{harris} catalog, and the same values
adopted there for the average stellar mass ($0.3 M_\odot$) and
$V$-band mass-to-light ratio ($M/L_V = 2$), the new determinations of
the structural parameters, distance, and extinction quoted in Table
\ref{tab_params} bring to $\log(t_{rc})=7.4$ and $\log(t_{rh})=9.0$
(in units of year). If the effective radius is used in place of $r_h$
(as it is done, e.g., in the Harris catalog and in
\citealp{mclaughlin+05}), the relaxation time becomes smaller than 1
Gyr: we find $\log(t_{re})=8.8$, which is 37\% shorter than $t_{rh}$.
For comparison, the central relaxation time quoted in the
\citet{harris} catalog is $\log(t_{rc})=7.6$, i.e., a factor of $\sim
1.6$ longer than our determination, and the median relaxation time
(8.62 in logarithimic units) is a factor of $\sim 1.5$ shorter than
our value of $t_{re}$, mainly reflecting the scale-length differences
discussed above, while the assumption of different extinction law and
color excess has a negligible impact on the result. These values
suggest that NGC 6440 is in a dynamically evolved stage (its age being
much longer than the relaxation times), although the $A^+$ parameter
has not been determined yet for this system, because a safe selection
of its blue straggler population has been hampered so far by the large
contamination from Galactic field stars and the severe and
differential reddening conditions along its line of sight.

\begin{figure*} 
\begin{center}
\includegraphics[width=150mm]{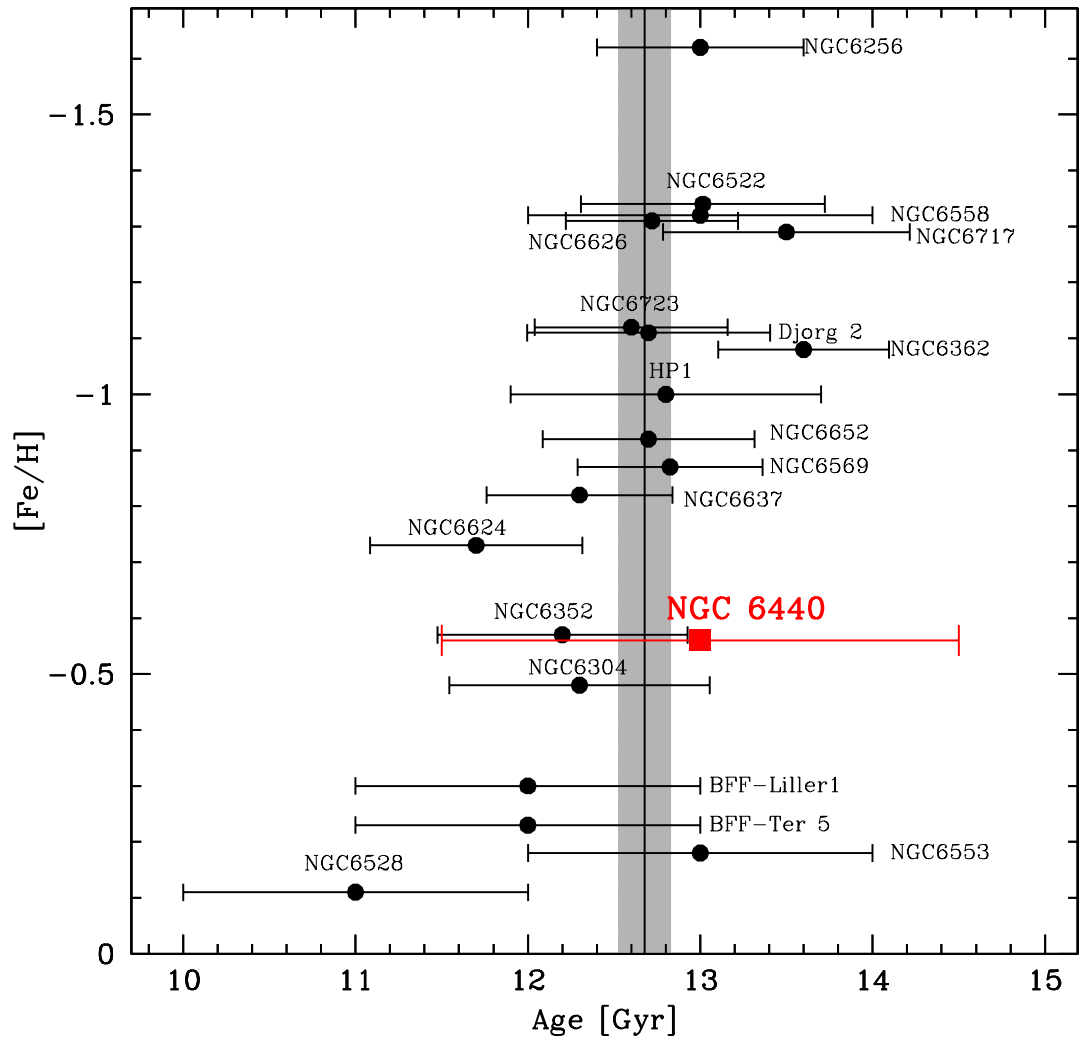}
\caption{ Age-metallicity distribution for the bulge GCs with age
  estimate available. NGC 6440 is marked as a large red square. 
  Data for the other clusters are mainly from 
    \citet[][see their Figure 16]{saracino+19}  and \citet[][see their Figure 12]{oliveira+20}  with the   
    addition of the recent age determination of NGC 6256 \citep{cadelano+20}. 
   We also plotted the age-metallicity of the oldest
stellar population in the two Bulge Fossil Fragments (BFF; namely Terzan5 and Liller1) so far
discovered into the bulge \citep{ferraro+09b,ferraro+16b,ferraro+21}. The grey vertical strip marks the
  weighted average and $1 \sigma$ uncertainty ($12.7 \pm 0.2$ Gyr) of
  the entire sample.}
\label{Fig:ageZ}
\end{center}
\end{figure*}

\begin{table}
\caption{Identity card of NGC 6440: new determinations of its
    basic parameters.}
\begin{center}
\begin{tabular}{ l l}
\hline
\hline
Parameter & Estimated value \\
\hline
Center of gravity & $\alpha_{\rm J2000}=17^{\rm h} 48^{\rm m} 52.84^{\rm s}$ \\
                  & $\delta_{\rm J2000} = -20^{\circ} 21\arcmin 37.5\arcsec$\\
Reddening law     & $R_V=2.7$\\
Color excess      & $E(B-V)=1.26-1.28$\\
Distance modulus  & $\mu_0=14.6\pm 0.1$\\
Distance          & $d=8.3\pm 0.4$ kpc\\
Age               & $t=13\pm 1.5$ Gyr\\
RGB bump & $M_{\rm V_{606}}^{Bump}=1.12\pm 0.12 $\\
&$M_{J}^{Bump}=-0.37\pm 0.12 $ \\
&$M_{K}^{Bump}=-0.94\pm 0.12 $ \\
Dimensionless central potential & $W_0=8.10^{0.20}_{-0.20}$\\
Concentration parameter & $c=1.86^{0.06}_{-0.06}$ \\
Core radius             & $r_c=6.4^{0.3}_{-0.3}$ arcsec $= 0.26^{0.01}_{-0.01}$ pc\\
Half-mass radius        & $r_h=50.2^{5.2}_{-4.5}$ arcsec $= 2.02^{0.21}_{-0.18}$ pc\\
Effective radius        & $r_e=36.8^{3.7}_{-3.2}$ arcsec $= 1.48^{0.15}_{-0.13}$ pc\\
Truncation radius       & $r_t=481.4^{43.9}_{-42.3}$ arcsec $= 19.4^{1.8}_{-1.7}$ pc\\
Central relaxation time & $\log(t_{rc}/{\rm yr}) = 7.4$\\
Half-mass relaxation time & $\log(t_{rh}/{\rm yr}) = 9.0$\\
\hline
\end{tabular}
\end{center}
\label{tab_params}
\end{table}

\acknowledgments  We thank the anonymous referee for useful
  comments that improved the presentation of the paper. This work is
part of the project {\it Cosmic-Lab} at the Physics and Astronomy
Department of the Bologna University
(\texttt{http://www.cosmic-lab.eu/Cosmic-Lab/Home.html}). The research
was funded by the MIUR throughout the PRIN-2017 grant awarded to the
project {\it Light-on-Dark} (PI:Ferraro) through contract
PRIN-2017K7REXT.  The research is based on observations collected with
the NASA/ESA HST (Prop. 11685, 12517, 13410 and 15403), obtained at
the Space Telescope Science Institute, which is operated by AURA,
Inc., under NASA contract NAS5-26555.  LC is the recipient of the ARC
Future Fellowship FT160100402.  SS gratefully acknowledges financial
support from the European Research Council (ERC-CoG-646928,
Multi-Pop).

\vspace{5mm}
facilities:{HST(WFC3/UVIS)};ESO(FORS2)



\end{document}